\documentclass[twoside,slac_one]{revtex4}
\usepackage{graphicx}
\usepackage{fancyhdr}
\usepackage{amsmath} 
\usepackage{bm}
\usepackage{amsxtra}
\usepackage{amssymb}
\usepackage{amsthm}
\usepackage{latexsym}
\usepackage{lscape}

\newcommand{\ptsj}{p_T^{\rm jet2}}
\newcommand{\gpTWOj}{$\gamma+{\rm 2~jet}$}
\newcommand{\gpTHRj}{$\gamma+{\rm 3~jet}$}
\newcommand{\Pt}{\ensuremath{p_T}}

\newcommand{\DO}{D\O{}}

\newcommand{\DPhi}{\Delta\phi}

\begin{document}

\title{Studies of multi-parton interactions in photon+jets events at D0}

%

\author{Dmitry Bandurin ~({\it for the DO Collaboration})}
\affiliation{Department of Physics, Florida State University, Tallahassee, FL, USA}

\begin{abstract}
      We consider sample of inclusive \gpTHRj~ events collected by the D0 experiment.
      The double parton fraction ($f_{\rm DP}$) and effective cross section $\sigma_{\rm eff}$,
      a process-independent scale parameter related to the parton density
      inside the nucleon, are measured in three intervals
      of the second (ordered in \Pt) jet transverse momentum $\ptsj$
      within the $15 \leq \ptsj \leq 30$ GeV range.
     Also we measured cross sections
     as a function of the angle in the plane transverse to the beam direction
     between the transverse momentum (\Pt)
     of the $\gamma+$leading jet system
     and \Pt~ of the other jet for \gpTWOj,
     or \Pt~ sum of the two other jets for \gpTHRj~ events.
     The results are compared to different models of multiple parton interactions (MPI)
     in the {\sc pythia} and {\sc sherpa} Monte Carlo (MC) generators.
\end{abstract}

\maketitle

\thispagestyle{fancy}

\section{Introduction}

  Many features of high energy inelastic hadron collisions
  depend directly on the parton structure of hadrons.
  The inelastic scattering of nucleons occurs mainly through 
  a single parton-parton interaction but the contribution from 
double (or multiple) parton collisions can be significant.
Information about DP rates is needed for understanding of 
nature of MPI events and correct estimating  background 
to many rare processes, especially with multi-jet final state.

\section{Double parton interactions in $\gamma$+3 jet events} 

The cross section of DP events production directly proportional
to cross sections of two processes A and B and
should be normalized by some scaling parameter $\sigma_{\rm eff}$ in cross section's units.
  \begin{eqnarray}
   \sigma_{DP} \equiv
     \frac{\sigma^{A} \sigma^{B}}{\sigma_{\rm eff} }.
   \label{eq:sigma_DPS}
   \end{eqnarray}
In general sense, $\sigma_{\rm eff}$ is a factor which characterizes
a size of the effective interaction region.

We have used a sample of $\gamma + 3$ jets events collected by the DO experiment
with an integrated luminosity of about 1~fb$^{-1}$.
The D0 detector is a general purpose detector
described in \cite{D0_det}.
The events should pass triggers based on the identification
of high $p_T$ cluster in the EM calorimeter with loose shower shape 
requirements for photons.
Jets are reconstructed using
the D0 Run~II iterative midpoint cone algorithm \cite{Run2Cone}
with a cone size $0.7$. 
Each event must contain
at least one $\gamma$ in the rapidity 
region $|y|<1.0$ or $1.5<|y|<2.5$
and at least three jets with $|y|<3.0$. 
Events are selected with $\gamma$ transverse momentum 
$60<p^{\gamma}_{T}<80$ GeV, leading (in $p_T$) jet $p_T>25$ GeV, while
the next-to-leading (second) and third jets must have $p_T>15$ GeV.
The DP fractions and $\sigma_{\rm eff}$ are determined
in three $\ptsj$ bins: 
15--20, 20--25, and 25--30 GeV.

\begin{figure}[h]
\vspace*{-7mm}
\hspace*{4mm} \includegraphics[scale=0.29]{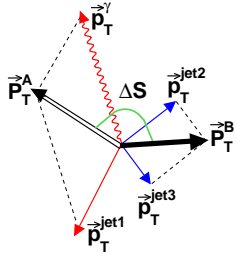}
~\\[-9mm]
\caption{A possible orientation of photon and jets transverse
momenta vectors in \gpTHRj~ events. Vectors $\vec{P}_{T}^1$ and $\vec{P}_{T}^2$
are the $p_T$ imbalance vectors of $\gamma$+jet and jet-jet pairs. The
figure illustrates a general case for the production of $\gamma$+3 jets +X events.}
\label{fig:dS}
\end{figure}

We use rates of double interactions (DI) from two separate $p\bar{p}$ 
collisions and DP from a single $p\bar{p}$ collision 
to extract $\sigma_{\rm eff}$ from their ratio \cite{prd1}.
The DI events differ from the DP events 
by the fact that the second parton scattering happens at a separate $p\bar{p}$ collision vertex.
Data events with a single $p\bar{p}$ collision vertex, which compose the sample of DP candidates, 
are selected separately from events with two vertices which compose
a sample of DI candidates. A distinctive feature of the DP events is 
a presence of two independent parton-parton scatterings 
within the same $p\bar{p}$ collision. 
We define the variable sensitive to the kinematics of the DP events:
\begin{eqnarray}
\Delta S \equiv \Delta\phi\left(\vec{p}_{T}^{\gamma,jet1}, ~\vec{p}_{T}^{jet2,jet3}\right),
\label{eq:DeltaSvar}
\end{eqnarray}
where
$\Delta\phi$ is an azimuthal angle between the $p_T$ vectors of
the total transverse momenta of the two two-body systems,
$\vec{p}_{T}^{\gamma, jet1}$ and $\vec{p}_{T}^{jet2, jet3}$, in \gpTHRj~ events.
This angle is schematically shown in Fig.~\ref{fig:dS}.

\begin{figure}[h]
\vspace*{-5mm}
\hspace*{4mm} \includegraphics[scale=0.32]{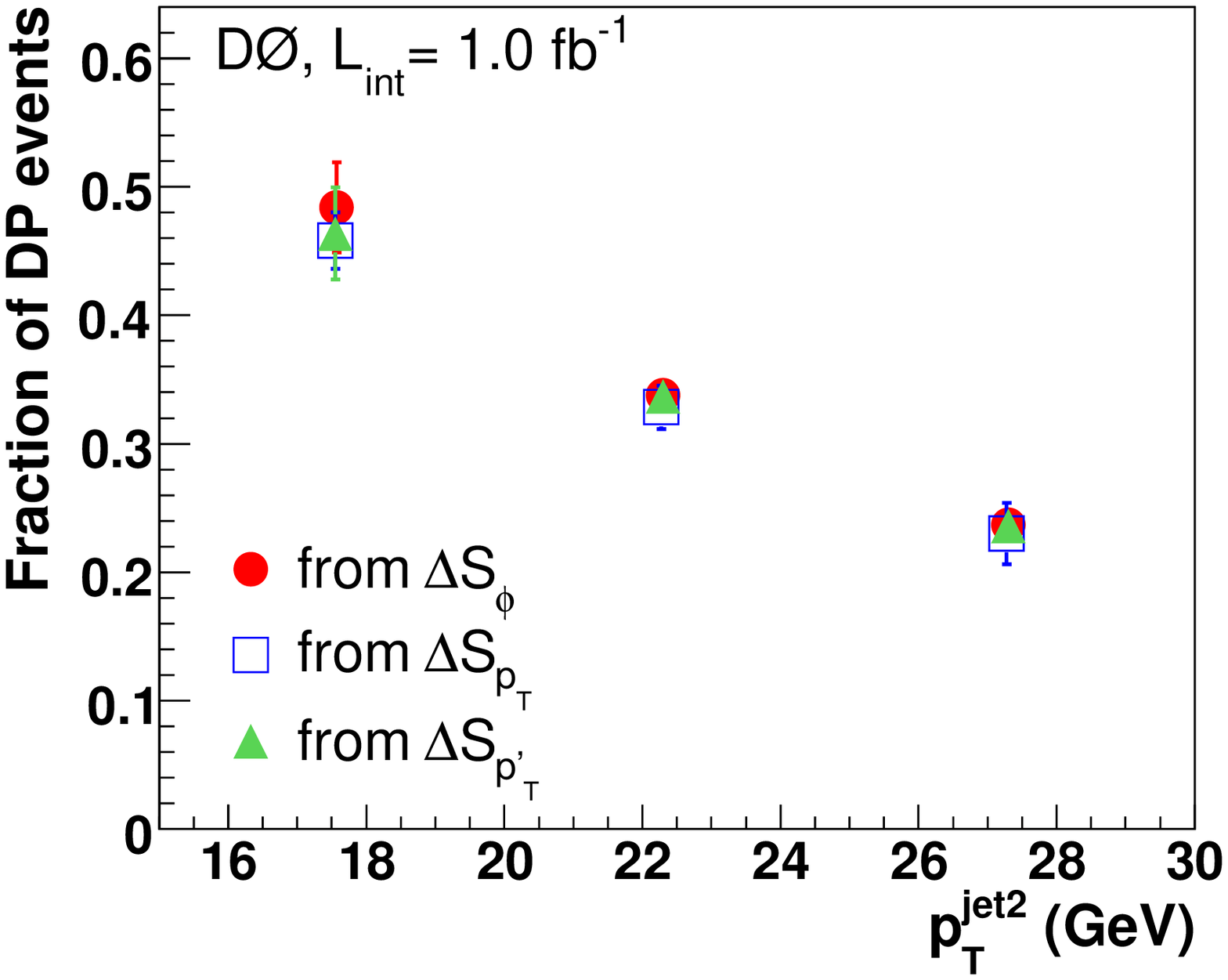}
~\\[-3mm]
\caption{Fractions of \gpTHRj events with double parton interactions in the three $\ptsj$ intervals.}
\label{fig:dpfrac}
\end{figure}

\begin{figure}[h]
\vspace*{-7mm}
\hspace*{4mm} \includegraphics[scale=0.32]{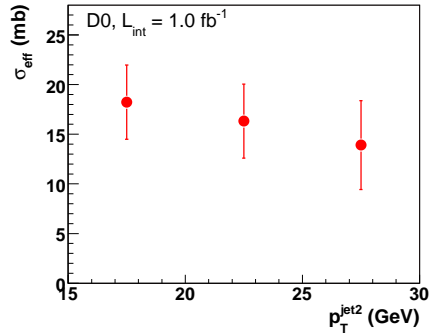}
~\\[-3mm]
\caption{Effective cross section $\sigma_{\rm eff}$ (mb) measured in the three $\ptsj$ intervals.}
\label{fig:sigma_eff}
\end{figure}

We consider the data-driven method to extract the DP fractions $f_{\rm DP}$.
Specifically, we consider data 
in two adjacent $\Pt$ intervals of the second jet.
The distribution for $\Delta S$ variable in data can be expressed as a sum of signal and background distributions.
If we known properties of data and DP model, 
the only unknown parameter is the fraction of DP events in one $\ptsj$ bin.
It is obtained from a minimization.
The found $f_{\rm DP}$ values with total uncertainties are
$0.466\pm 0.041$ for $15<\ptsj <20$ GeV, $0.334 \pm 0.023$ for $20<\ptsj<25$ GeV,
and $0.235 \pm 0.027$ for $25< \ptsj <30$ GeV. They are shown in Fig.~\ref{fig:dpfrac}
(three sets of the points correspond to three possible definitions for the $\Delta S$ variable \cite{prd1}).
The values of $\sigma_{\rm eff}$~ are shown in Fig.~\ref{fig:sigma_eff}.
The main systematic uncertainty are caused by determinations of the DI and 
DP fractions giving a total systematic uncertainty of $(20.5-32.2)\%$.
The obtained $\sigma_{\rm eff}$ values in different $\ptsj$ bins
agree with each other within their uncertainties and highly uncorrelated,
and are used to calculate the average value:
\begin{equation}
 \sigma_{\rm eff}^{\rm ave} = 16.4 \pm 0.3(\rm stat) \pm 2.3(\rm syst)  {~~\rm mb}. 
  \label{eq:sigeff_av}
\end{equation}
      This average value is
      in the range of those found in previous measurements \cite{AFS,UA2,CDF93,CDF97}
      performed at different energy scales of parton interactions.

\section{Azimuthal decorrelations and multiple parton
  interactions in ~\gpTWOj~ and ~\gpTHRj~ events in $p\bar{p}$ collisions}

\begin{figure}[h]
\vspace*{-7mm}
\hspace*{4mm} \includegraphics[scale=0.27]{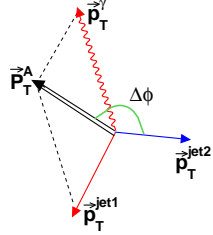}
~\\[-9mm]
\caption{A possible orientation of photon and jets transverse
momenta vectors in \gpTWOj~ events.}
\label{fig:dPHI}
\end{figure}

\begin{figure}[htbp]
\hspace*{0mm}  \includegraphics[scale=0.32]{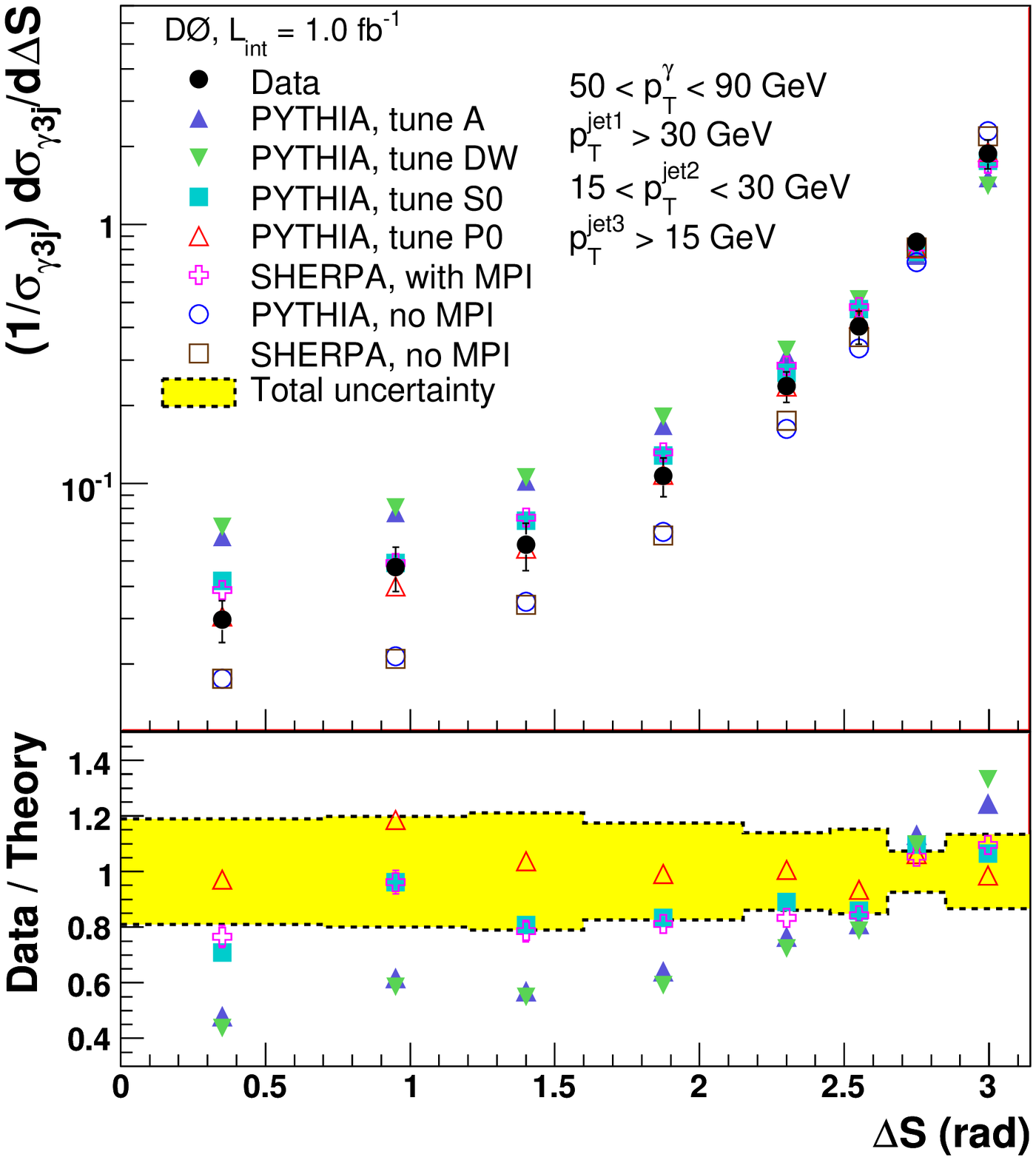}
\hspace*{5mm}  \includegraphics[scale=0.32]{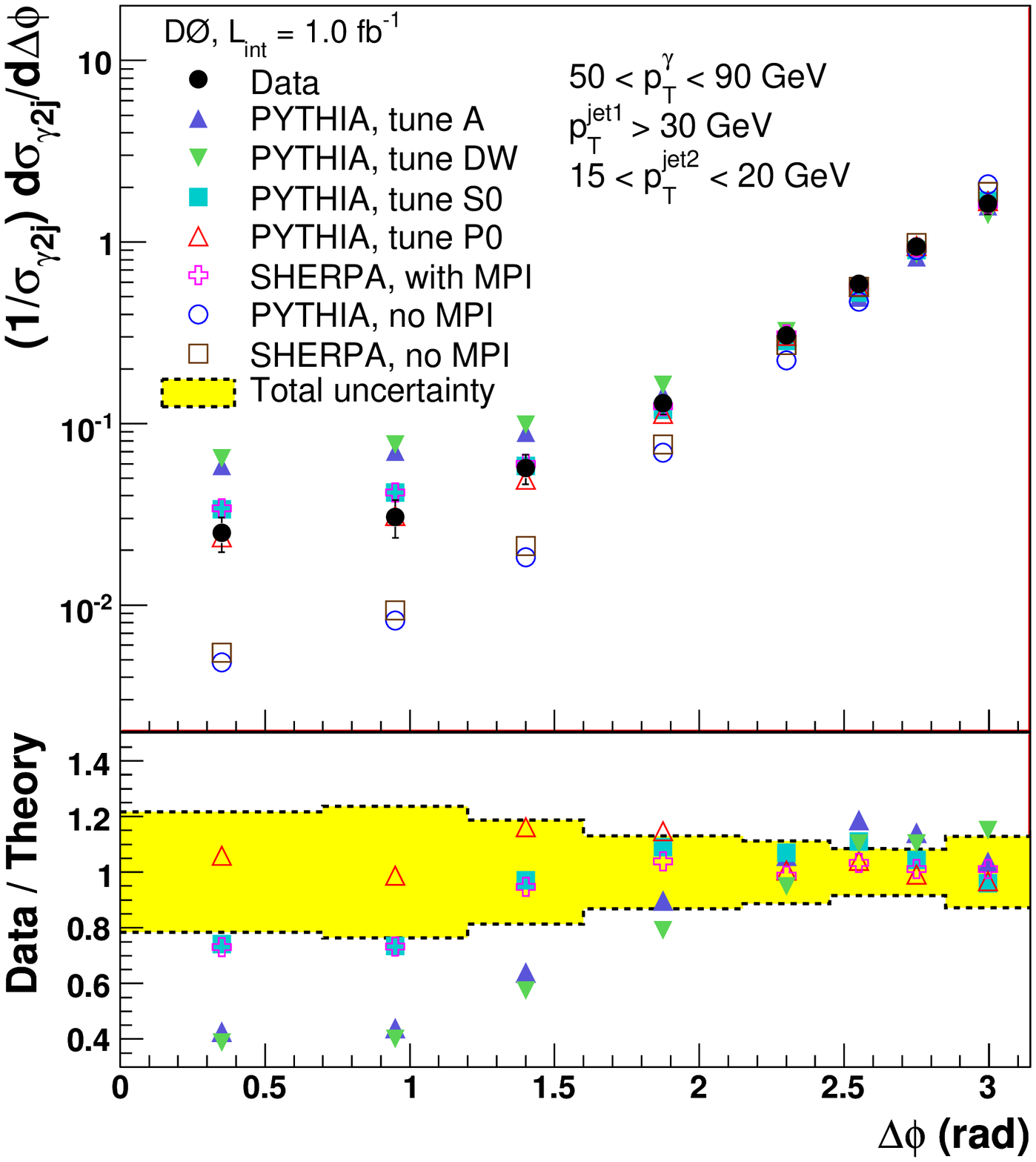}
~\\[-2mm]
\caption{Left: Normalized differential cross section in the $\gamma + 3$-jet
events, $(1/\sigma_{\gamma 3j})\sigma_{\gamma 3j} /d\Delta S$, in data compared to MC models
and the ratio of data over theory, only for models including
MPI, in the range $15 < p_T^{jet2} < 30$ GeV.
Right: Normalized differential cross section in $\gamma + 2$-jet
events, $(1/\sigma_{\gamma 2j})\sigma_{\gamma 2j} /d\Delta\phi$, in data compared to MC models
and the ratio of data over theory, only for models including
MPI, in the range $15 < p_T^{jet2} < 20$ GeV.}
\label{fig:ds_d0}
\end{figure}

Samples of $\gamma + 2(3)$ jets events with the same cuts as \cite{prd1} are considered.
The next modifications are applied: each event must contain at least one $\gamma$ in the pseudorapidity
region $|y|<1.0$ or $1.5<|y|<2.5$
and at least two (or three) jets with $|y|<3.5$.
Events are selected with $\gamma$ transverse momentum
$50<p^{\gamma}_{T}<90$~GeV, leading jet $p_T>30$~GeV, and
the second jet $p_T>15$~GeV.
If there is a third jet with $p_T>15$ GeV that passes the selection criteria, the event
is also considered for the $\gamma + 3$ jet analysis.
By measuring differential cross sections vs. the azimuthal angles 
in $\gamma + 2(3)$ jet events we can better tune MPI models in events with high \Pt~ jets.
We present the four measurements of normalized differential cross sections, 
 $\Delta S$ in a single bin $15<\ptsj<30$ for ~\gpTHRj~ events 
(see Fig.~\ref{fig:dS}), and
 $\Delta\phi$ in three $\ptsj$  bins, 15-20, 20-25, and 25-30 GeV, for \gpTWOj~ events.
The $\Delta\phi$ is an angle between the $p_{T}$ vector obtained by pairing the $\gamma$ and the leading jet $p_T$ vectors 
and the second jet $p_T$ vector ~\cite{prd2}). It is shown in  Fig.~\ref{fig:dPHI}.
We consider a few MPI models and two models without MPI simulated by {\sc pythia} and {\sc sherpa} MC generators.
Figure~\ref{fig:ds_d0} shows the measured cross section for
the two angular variables $\Delta S$ (left plot) and $\Delta\phi$ (right plot).
The data have a good sensitivity  to the various MPI models, which predictions 
vary significantly and differ from each other by up to a factor 2 at small $\Delta S$ and 
$\Delta\phi$, i.e. in the region where the relative DP contribution is expected to be highest.

From these two plots we may conclude that: (a) a large difference between single parton-parton interaction (SP) models 
and data confirms a presence of DP events in the data sample;
(b) the data favor the predictions of the MPI models with P0, S0 and Sherpa MPI tunes with $p_T$-ordered showers;
(c) the predictions from tune A and DW MPI models are disfavored.
It is important that our preferable choice of MPI models is stable for all our measurements. 

\begin{figure}[htbp]
\hspace*{0mm}\includegraphics[scale=0.34]{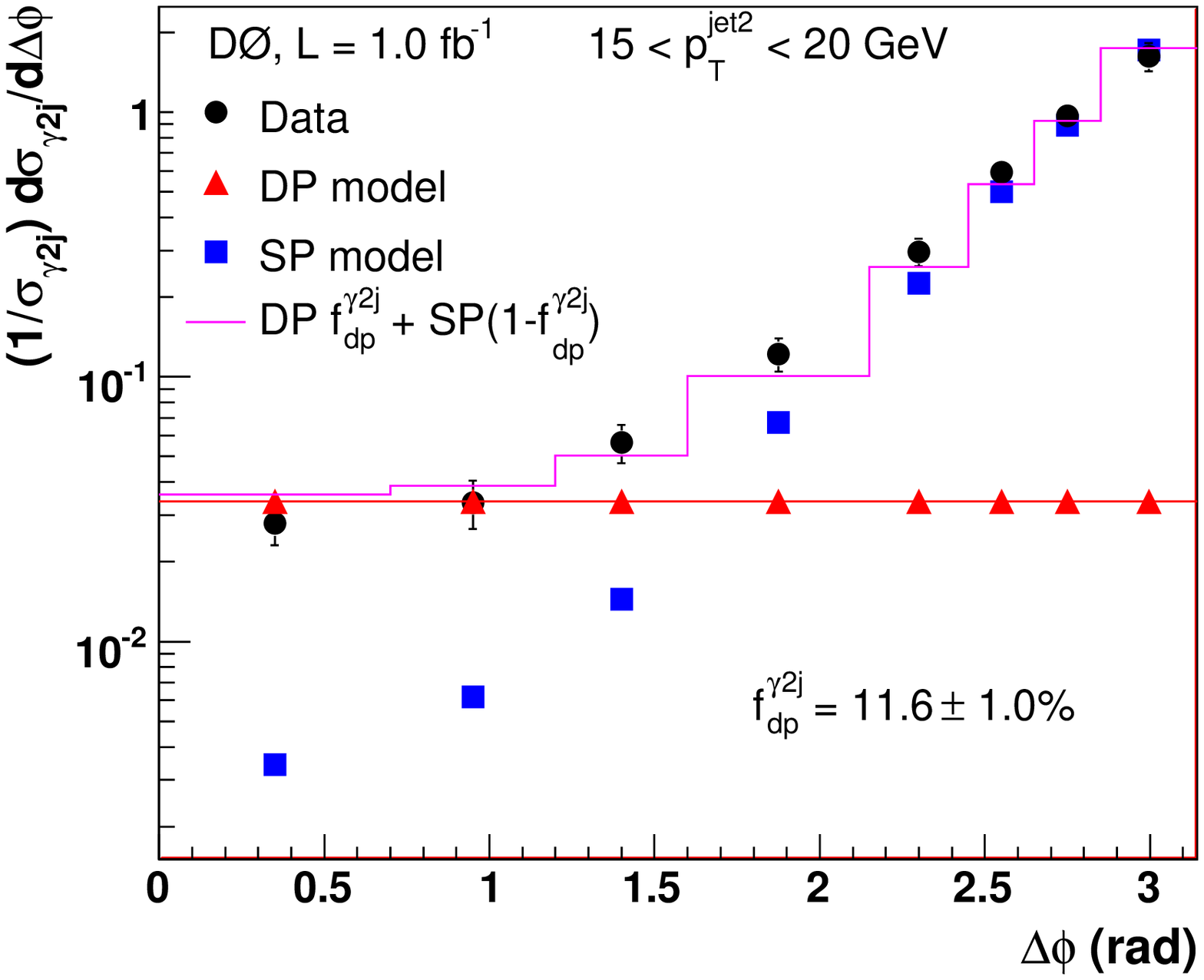}
\hspace*{5mm}\includegraphics[scale=0.34]{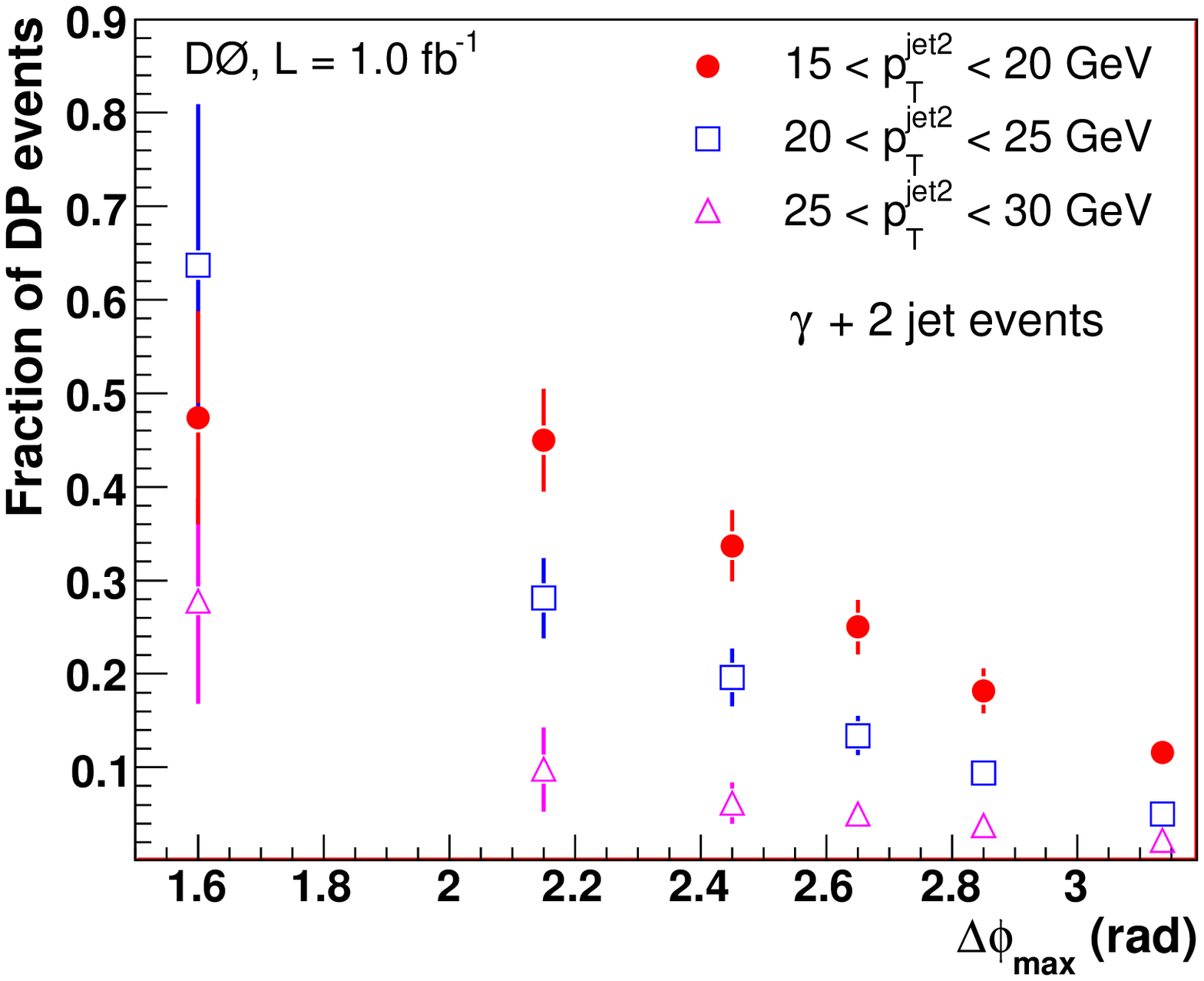}
~\\[-3mm]
\caption{Left: the $\Delta \phi$ distribution in data, SP, and DP models, and the sum of the SP and DP contributions
weighted with their fractions for $15<\ptsj<20$ GeV.
Right: the fractions of DP events with total uncertainties in \gpTWOj final state as a function of
the upper limit on $\DPhi$ for the three $\ptsj$ intervals.}
\label{fig:fract}
\end{figure}

In \gpTWOj~ events in which the second jet is produced in the additional independent parton interaction, 
the $\Delta\phi$ distribution should be flat.
Using this fact and also SP prediction for $\Delta\phi$ we can get the DP fractions  from a fit to data.
The distributions in data, SP, and DP models, as well as a sum of the SP and DP distributions,
weighted with their respective fractions for $15<\ptsj<20$ GeV, are shown in the left plot of Figure~\ref{fig:fract}.
The DP fractions in the \gpTWOj~ samples decrease in the bins of $\ptsj$ as
$(11.6\pm1.0)\%$ for $15-20$ GeV, $(5.0\pm1.2)\%$ for $20-25$ GeV,
and $(2.2 \pm 0.8)\%$ for $25-30$ GeV.
To determine the fractions as a function of $\Delta\phi$,
we perform a fit in the different $\Delta\phi$ regions by excluding the bins at high $\Delta\phi$. 
We find that they grow significantly towards the smaller angles and are higher for 
smaller $\ptsj$ (right plot of Figure~\ref{fig:fract}).

\begin{figure}[h]
\vspace*{-5mm}
\hspace*{4mm} \includegraphics[scale=0.32]{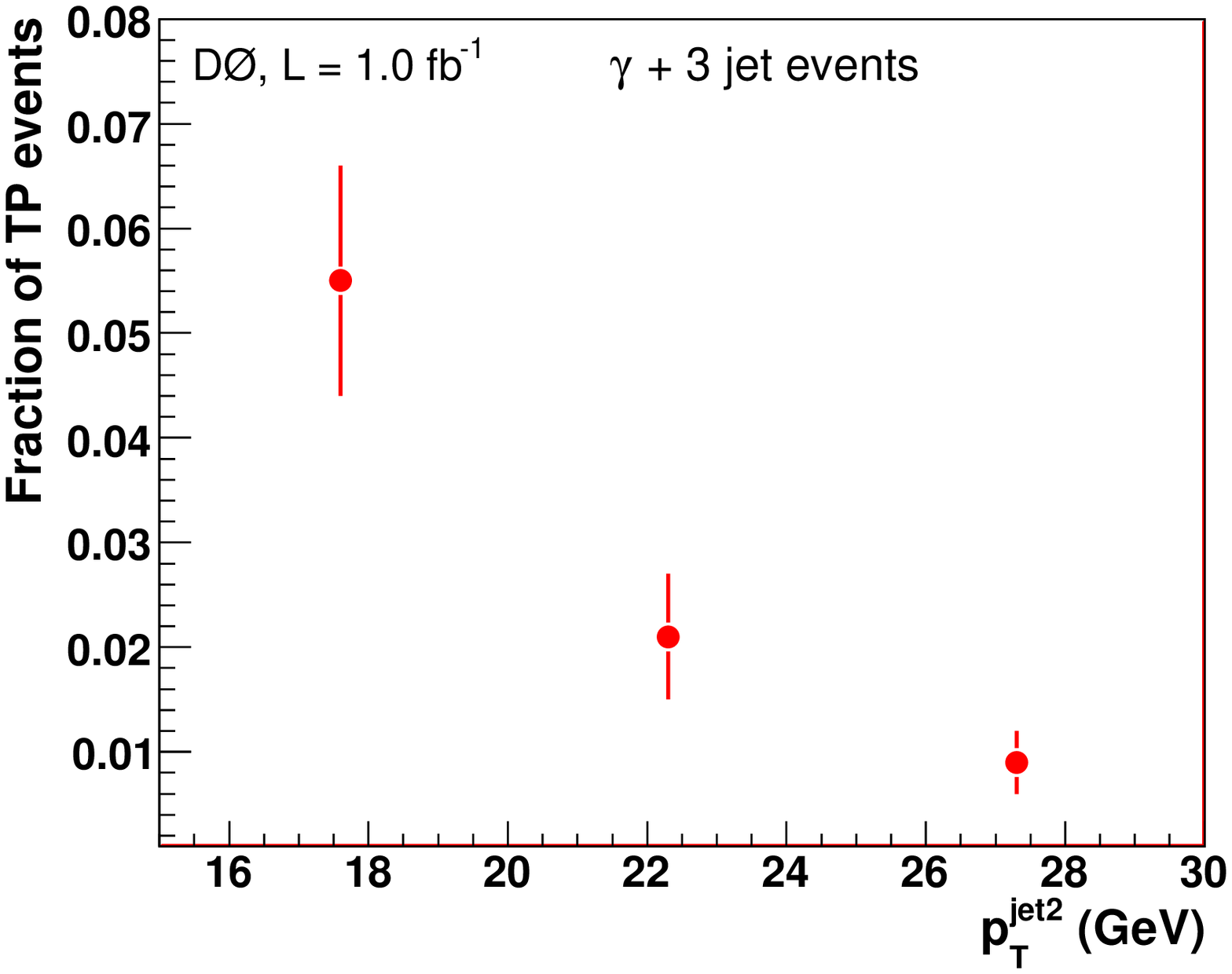}
~\\[-3mm]
\caption{Fractions of \gpTHRj events with triple parton interactions in the three $\ptsj$ intervals.}
\label{fig:tpfrac}
\end{figure}

We also estimate the fraction of \gpTHRj~ events from
triple parton interactions (TP) in data as a function of $\ptsj$.
In \gpTHRj~ TP events, the three jets come from three different
parton interactions, one $\gamma +$ jet and two dijet final states.
In each of the two dijet events, one of the jets is either not reconstructed or
below the 15 GeV $p_T$ selection threshold.
The fractions of TP events in the \gpTHRj~ data have been estimated and are shown in Fig.~\ref{fig:tpfrac}.
As we see, they vary in the $\ptsj$ bins as $(5.5\pm1.1)\%$ for $15-20$ GeV, $(2.1\pm0.6)\%$
for $20-25$ GeV, and $(0.9 \pm 0.3)\%$ for $25-30$ GeV.





\end{document}